\documentstyle[aps,epsf]{revtex}

\begin{document}

\title{Temporal fractal structures: Origin of power-laws in the world-wide Web}

\author{Bosiljka Tadi\'c$^*$}

\address{Jo\v{z}ef Stefan Institute,
P.O. Box 3000, 1001 Ljubljana, Slovenia}

\maketitle

\begin{abstract}
Using numerical simulations and scaling theory we study the
dynamics of the world-wide Web from the growth rules recently proposed in
Ref.\ [1] with appropriate parameters. We demonstrate that the
emergence of power-law behavior of the out- and in-degree distributions
in the Web involves occurrence of temporal fractal
structures, that are  manifested in  the
scale-free growth of the local connectivity and in first-return time
statistics. We also show how the scale-free behavior occurs in the
statistics of random walks on the Web, where the walkers use information
on the local graph connectivity.
\end{abstract}
\pacs{PACS numbers: 05.40.Fb, 89.20.Hh, 89.75.Da}

\section{Introduction}
Complex evolving networks differ from static random graphs in that
their size increases in time, thus impacting the linking process in a
nontrivial manner. Hence the {\it emergent} structure of links is related
to salient features of the growth processes in the network, which
is governed  by the individual linking rules \cite{Strogatz,Reka}.
In the case of the scale-free structures emerging at large evolution
times in networks with {\it preferential} linking, the universality
classes characterized by the same set of scaling exponents can be
distinguished, that are based on several {\it  relevant details} of the
microscopic linking properties. The theoretical background of the
universality classes of dynamic networks is still missing. By study of
many particular networks it has been recognized that  certain
 dynamic constraints on the linking processes
can change the  emergent scale-free behavior~\cite{BT1,Strogatz,Reka}.

The world-wide Web differs from the generic scale-free evolving
networks in two ways: (1) it is represented by a {\it directed graph} and,
(2) most importantly, it has a {\it variable wiring diagram}.
Frequent updates of the out-going links, that are peculiar for
 conduct of the agents in the real Web, makes the wiring diagram of the
Web graph changing at the same paste as the graph grows. Whereas, wiring
diagram of some other networks  changes on much slower scale or not at
all~\cite{BT1}. The intimate relationship between structural and growth
properties leads to a specific architecture of links in the Web. Recent
measurements in the real Web have shown~\cite{WWW} that both out- and
in-degree distributions are power-law with {\it different} exponents, as
well as the size of the connected clusters  out of the giant component.

Occurrence of power-laws is a remarkable feature in large number of complex
evolving networks, that indicates presence of underlying self-organization
while the network grows. The emergent hierarchical organization of node ranks
is
highly relevant for the stability of the network under attacks \cite{attack},
and  for the character of other dynamic processes {\it on} the network, such as
random walk processes \cite{BT2,BT3}. Therefore understanding the mechanisms
of self-organization that lead to power-laws in the world-wide Web is
crucial both for its functional stability and for
designing efficient search algorithms \cite{algorithms} and
transport processes \cite{transport} on the Web graph.

As a step towards realistic modeling of the dynamics of world-wide Web
we proposed recently the model~\cite{BT1} which takes into account the
basic relevant features of the Web growth: directed linking, rewiring
of preexisting links, and bias activity of agents and bias attachment of
links. It was shown~\cite{BT1} that, when the degree of rewiring in the
graph is adjusted to $\beta \approx 3$ (i.e., to each new added link in the
graph there are in the average three updated links among preexisting nodes),
the model reproduces fairly well the emergent power-law distributions
of the out- and in-degree, and the scaling exponent of the connected
components.  In the present work we use this model to study the details
of the growth process that precedes the emergence of link structure in the
Web graph. Particularly, we demonstrate that a spatio-temporal {\it fractal
structure of linking activity} occurs on the growth time scale by successive
addition of nodes and the average increase of the number of links by
$M=\beta +1$. The fractal properties of the structure are measured by
scaling behavior of the distribution
of time intervals  of the {\it first-return} activity to a given node.
We show that this activity pattern results in the algebraic increase of
the average  {\it local  connectivity} $<q_\kappa (s,t)>$ at node $s$
with time ($\kappa $ refers to ``out'' and ``in'' links), implying scaling
behavior in the underlying local probability distribution.
In view of the  scaling theory the local connectivity is then
related  to the emergent degree distribution in the
limit of large evolution times~\cite{DMS} $t \to \infty $.
We demonstrate these steps by directly simulating the appropriate
quantities, both for out- and in-links. In addition, we show that
dynamic processes, such as
random walks on the Web graph~\cite{BT2,BT3}, that use the information on
the local connectivity may also result in the power-law distributions.
Here we compute (for the same graph parameters) the distributions of
distances on node hierarchy made by an ensemble of random walkers which
utilize parts of locally available information on the Web structure.

\section{Linking Rules and Fractal Growth Patterns}

The growth model is defined by the dynamic rules~\cite{BT1} that can be
summarized as follows: At each time step $t> M_0\geq M$ add a node $i = t$
and create $M$ links. A link is first attempted from the new added node with
probability $\alpha $ to a target node $k$ that is selected with the
probability $p_{in}(k,t)$, specified below. Else, a link is created
between a pair of preexisting nodes  (updated link) as follows:
 A link from node $n< i$ to target node $k< i$ at time $t=i$
occurs with the probability
\begin{equation}
C(n,k,t) = (1-\alpha )\ p_{out}(n,t)\times p_{in}(k,t) \ ,
\label{cnk}
\end{equation}
where both probability to select an origin of the link  $p_{out}(n,t)$
and to select  a target node $p_{in}(k,t)$  depend on the
current connectivity of these nodes $q_{out}(n,t)$ and $q_{in}(k,t)$,
respectively,
\begin{equation}
p_{out}(n,t) = {{\alpha + q_{out}(n,t)/M}\over{(1+\alpha )t}} \ , ~~~~
p_{in}(k,t) = {{\alpha + q_{in}(k,t)/M}\over{(1+\alpha )t}}  \ .
\label{pp}
\end{equation}
At the moment of addition $q_{out}(i,i)=q_{in}(i,i) = 0$ and
increasing in time. Therefore, a ratio of the number of added and updated
links $\beta \equiv (1-\alpha )/\alpha $, which is independent on the actual
number of links $M$, is the control parameter in the model. For simplicity
we keep $M$ fixed, assuming that the number of links fluctuate in time
around the average value $M$.
Motivation for the above linking rules are discussed in detail~\cite{BT1}.
For $\beta \approx 3$ the results of numerical simulations within this model
 (see Ref.~\cite{BT1}) agree satisfactorily with the empirical data on
the real Web~\cite{WWW}.
We would like to stress that the property of rewiring while the graph
grows, which is enabled by $C(n,k,t) >0$ in Eq.\ (\ref{cnk}), yields
qualitatively new scaling features, compared with the graphs with frozen
links ($\beta \equiv 0$). For instance, one of the important consequences
of rewiring is the appearance of the scale-free structure of the out-degree
distribution. The entire class of graphs generated with the above rules
for varying $\beta $ in the range $0 < \beta <\infty $  was
studied recently~\cite{BT3}.

Here we concentrate to the growth phase of the Web graph. We fix $\beta =3$
and chose $M=4$. In principle, share between added and rewired links is
statistical, thus the universal degree-distributions are independent on
$M$ in the scaling region~\cite{DMS,BT1}. However, some local properties
can depend on the actual growth  rate of the number of links. To show
these dependences is another goal of this work.
\begin{figure}
\hskip 2 true cm
\epsfxsize=92mm \epsffile[34 36 456 266]{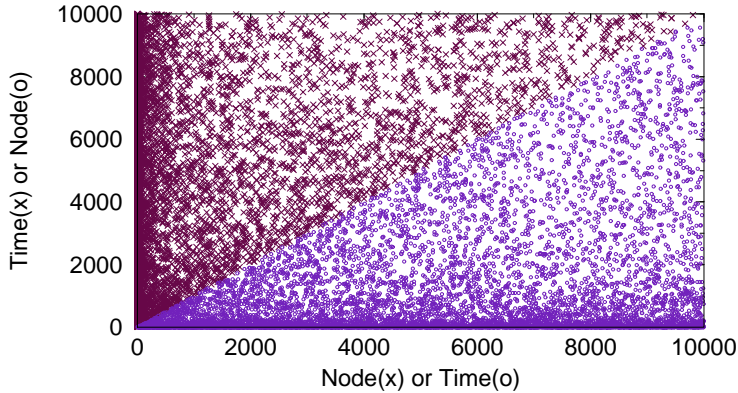}
\vskip 4mm
\caption{ \label{fig1}Temporal patterns of activity: in-linking (crosses) and
out-linking (circles) for $M=4$ and  network size $N=t=10^4$.}
\end{figure}
In Fig.\ 1 we show the activity pattern of each node during the growth time
$t=1 \cdots N$, with $N=10^4$ nodes-steps. A node $n$ can be linked only in the
time after its appearance $t\geq n$.  Value on the vertical axis of
a point in the plain $(k,t)$
represents the moment of time $t$ when the node $k$ got an in-link (upper-left
part), or a node $n$ that created an  out-link at time $t$ (lower-right part).
Quantitative characterization of these patterns can be done in two ways.

First, we notice that the time intervals between two consecutive linking
activities {\it at a given node} are irregular, resembling a fractal set.
For this reason we measure the distribution of successive time intervals
$\Delta t$ for in-linking and out-linking separately. The results are
given   in Fig. \ 2 (left), where the algebraic decay of the distributions
$P_\kappa (\Delta t) \sim (\Delta t)^{-\mu _\kappa }$, confirms the fractal
character of these sets. The slopes of the two curves are
$\mu _{out} = 0.82\pm 0.01$ and $\mu _{in} = 0.87\pm 0.01$.
Second, fixing a node $s$ we watch how the number of links accumulates at
that node with time. Averaging over a large ensembles of networks, we find
that  for $t\gg s$
\begin{equation}
<q_\kappa (s,t)> \sim  t^{\gamma _\kappa } \ ,
\label{avq}
\end{equation}
 where $\kappa \equiv $ ``out'' or ``in'', as shown in Fig.\ 2 (right), where
the scaling exponents are $\gamma _{out} = 0.66\pm 0.03$ and
$\gamma _{in} = 0.87\pm 0.03$.

\section{Local Connectivity and Emergent Degree Distributions}

 The average connectivity at a node increases as a power
of evolution time for times $t\gg s$, which is compatible with scaling
behavior of the local probability distribution
$\rho_\kappa(q,s,t)$ that the node $s$ collected $q$ links in time up to
the step $t$.  It was shown~\cite{DMS} analytically that preference
linking with the probability $p_{in}(k,t)$
given in Eq.\ (\ref{pp}) leads to the power-law behavior of the
$\rho _{in}(q,s,t)$ both in $q$ and $t/s$ arguments. Our results in Fig.\ 2
suggest that, due to rewiring with the probability $C(n,k,t)$ in Eq.\
(\ref{cnk}), the distribution $\rho _{out}(q,s,t)$ also exhibits scaling
behavior but with different  exponent ($\gamma _{out}\neq \gamma _{in}$). The
emergent degree distributions $P_{\kappa }(q)$ are defined as
\begin{equation}
P_\kappa (q) = \lim _{t\to \infty }\sum _{s < t}\rho _\kappa (q,s,t)
\sim q^{-\tau _\kappa } \ ,
\label{Pq}
\end{equation}
which we extend to both out- and in-links. In addition, the exact
 scaling relation that applies to in-link distributions~\cite{DMS} can be
easily extended to out-links \cite{TP}, i.e.,
\begin{equation}
\tau _\kappa = 1/\gamma _\kappa +1 \  .
\label{scaling-relation}
\end{equation}
Here we assume that the general scaling form applies both for in- and
out-links
\begin{equation}
\rho _\kappa (q,s,t) \sim
(s/t)^\mu f(q^x(s/t)^\Delta )\ ,
\label{scal-form}
\end{equation}
 with  conserved number of links
of both kind, i.e.,  $\sum _q \rho_\kappa (q,s,t) = 1$ where $\kappa =$ ``in'',
``out''. Then together with Eqs.\ (\ref{avq})-(\ref{Pq}) we find $\mu =\Delta
/x = \gamma $ and $\tau = (1+\gamma)x/\Delta $, leading to Eq.\
(\ref{scaling-relation}). The measured distributions of emergent
node ranks $P_{in}(q)$ and $P_{out}(q)$ after $N=10^5$ added nodes are shown
in Fig.\ 3. The slopes $\tau _{out}-1 =1.70\pm 0.03$ and $\tau _{in}-1=
1.26\pm 0.02$ obey the scaling relation (\ref{scaling-relation}) with
the respective values for $\gamma
_{out}$ and $\gamma _{in}$ taken from average connectivity in Fig.\ 2.
\begin{figure}[t]
\hskip 2 true cm
\epsfxsize=28pc
\epsfbox{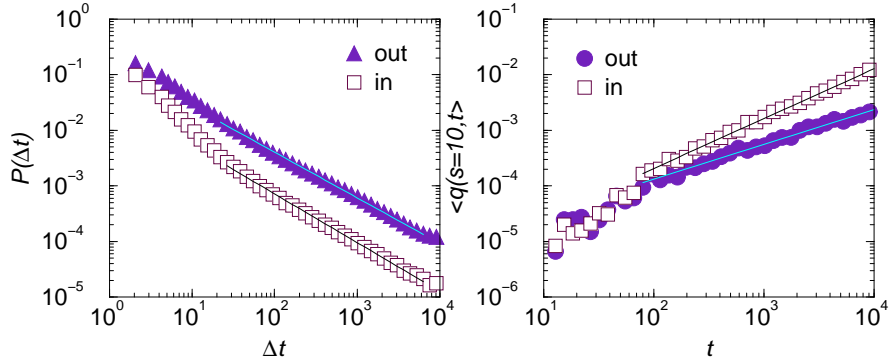}
\caption{ \label{fig2}Left panel: Distributions of return-time for out- and
in-linking. Right panel: Average connectivity at node $s=10$ vs. evolution
time $t$. $N=10^4$, $M=4$, data log-binned, bin ratio 1.2. }
\end{figure}

\begin{figure}[h]
\hskip 2 true cm
\epsfxsize=26pc
\epsfbox{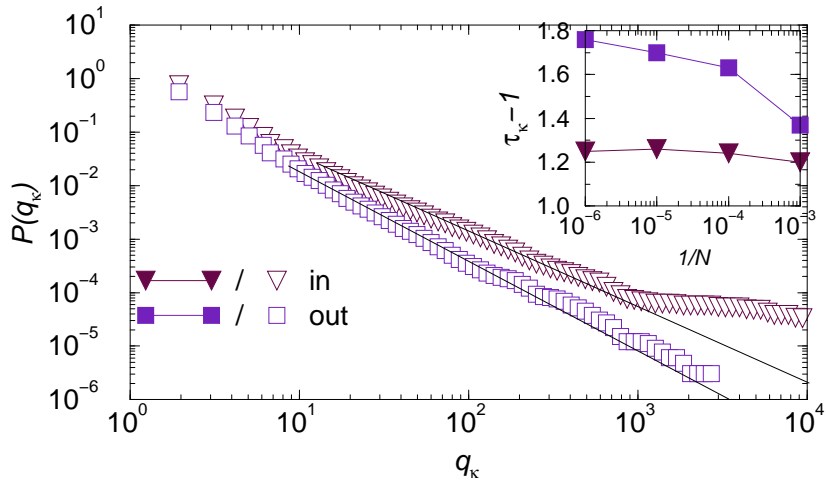}
\caption{ \label{fig3}Cumulative distributions of node degrees for out- and
in-links after $t=N=10^5$ evolution steps. $M=1$, log-binning ratio 1.1 .
Inset: Corresponding scaling exponents vs. $1/N$. }
\end{figure}

\section{Walker on the Web: Local Structures}
We have demonstrated that evolution of local organization  both of out- and
in-links at individual nodes is responsible for the global scaling of
the emergent node distributions for large times and number of nodes.
Here we would like to show that dynamic processes on grown network
(i.e., at different time scale) which use the information on these local
properties also obey certain scaling laws. Such processes on the Web are
different kinds of random walks
related, for example, to search algorithms. We define
two types of random walks~\cite{BT2,BT3}: The adaptive random walk (ARW)
that selects the target node $k_\ell $ with the weight which is proportional
to in-linking probability of the visited node, and a naive random walk (NRW)
selecting one of the out links with equal probability. The corresponding
 weights are
\begin{equation}
 w_{ARW}(n,k_\ell ) =  p_{in}(k_\ell)/\sum_{\ell =1}^{q_{out}(n)}
p_{in}(n,k_\ell )\  , \
 w_{NRW}(n,k_\ell ) = 1/q_{out}(n) \ .
\label{ww}
\end{equation}
In Fig.\ 4 we show the distributions of distances in hierarchy levels
$\Delta q_\kappa $ inside the clusters of {\it connected nodes} which are
visited in cumulative time by an ensemble of walkers. As the Fig.\ 4 shows,
these local clusters are organized scale-free structures  with
$  W(\Delta q_\kappa )\sim
(\Delta q_\kappa )^{-\delta _\kappa }$.
Notably the scaling exponents
$\delta _{out} \approx \delta _{in}$  expressing the correlations due to
normalization of weights in Eq.\ (\ref{ww}).
 For the case of ARW  the exponents are close to $\tau _{in} $
of the global structure of in-links, whereas for the clusters scanned by NRW
 they are reduced by approximately unity, i.e., $2.07\pm 0.04$ and
$1.10 \pm 0.03$ for ARW and NRW, respectively.

\begin{figure}[t]
\hskip 2 true cm
\epsfxsize=30pc e
\epsfbox{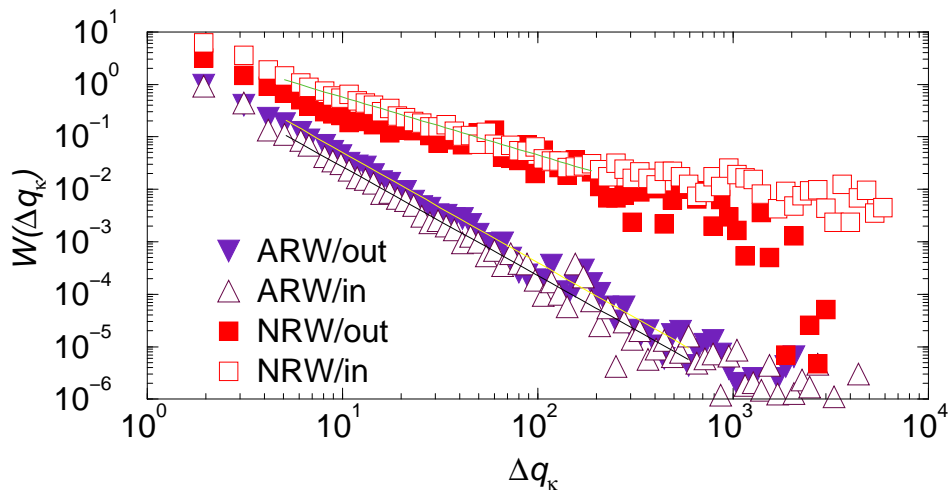}
\caption{ \label{fig4}Time-integrated distributions of distances of out- and
in-degree $\Delta q$, made by adaptive (ARW) and naive (NRW)  walkers on the
Web graph. $N=10^4$, $M=4$, log-binning ratio 1.1 .}
\end{figure}

\section{Conclusions}
The probabilistic character of linking rules for $\alpha < 1$ with
self-consistently
varying linking probabilities leads to power-law behavior of local and
global (emergent) link structures, both for out- and in-links.
 This  numerical results (see also  references~\cite{BT1,BT2,BT3})
 are in agreement with the analytical results  obtained by rate equation
approach in the same model \cite{TP}.
We have demonstrated here that the basis of these scaling laws lies in the
occurrence of dynamic fractals and hence the algebraic growth of the local
connectivity in time, from
which then the hierarchical global structure emerges at large times.

In addition, such local connectivity affects random-walk processes on grown
networks.
The connected clusters (subgraphs) scanned by the random-walk ensembles
also exhibit scaling behavior of distances in node degrees. The scaling
properties of these subgraphs on the Web strictly depend on the
applied random walk strategy,  i.e., in
 the degree of information about local connectivity that the walkers use.
Moreover, the scaling exponents of  the distributions of distances in
these subgraphs on the Web decrease with increasing rate $M$, in contrast
to the global structure of the graph, which is universal for large network
size $N$.

\section*{Acknowledgments}
This work  was  supported by the Ministry
of Education, Science and Sports of the Republic of Slovenia.

\end{document}